\documentclass{article}
\usepackage{hyperref}
\usepackage{amsmath}
\usepackage{amssymb}
\usepackage{graphicx}
\usepackage{authblk}
\usepackage{float}
\title{Graph Visualization for Blockchain Data}
\author[2]{Marcell Dietl}
\author[1]{Andre Gem\"und}
\author[1]{Daniel Oeltz}
\author[1]{Felix M. Thiele}
\author[2]{Christian Werner}

\affil[1]{Fraunhofer Institute for Algorithms and Scientific Computing SCAI, Schloss Birlinghoven, 53757 Sankt Augustin, Germany.}
\affil[2]{frontmark GmbH, Taunusstraße 63,
65183 Wiesbaden, Germany.}

\date{February 2024}

\usepackage{tikz}
\usetikzlibrary{shapes.geometric, arrows}

\tikzstyle{startstop} = [rectangle, rounded corners, 
minimum width=3cm, 
minimum height=1cm,
text centered, 
draw=black, 
fill=red!30]

\tikzstyle{process} = [rectangle, 
minimum width=3cm, 
minimum height=1cm, 
text centered, 
text width=3cm, 
draw=black, 
fill=orange!30]

\tikzstyle{decision} = [trapezium, 
trapezium stretches=true, 
trapezium left angle=70, 
trapezium right angle=110, 
minimum width=3cm, 
minimum height=1cm, text centered, 
draw=black, fill=blue!30]

\tikzstyle{arrow} = [thick,->,>=stealth]

\begin{document}

\maketitle
\begin{abstract}
    In this report, we introduce a novel approach to visualize extremely large graphs efficiently. Our method combines two force-directed algorithms, Kamada-Kawai and ForceAtlas2, to handle different graph components based on their node count. Additionally, we suggest utilizing the Fast Multipole method to enhance the speed of ForceAtlas2. Although initially designed for analyzing bitcoin transaction graphs, for which we present results here, this algorithm can also be applied to other crypto currency transaction graphs or graphs from diverse domains.
\end{abstract}
\section{Introduction}
Blockchain technology is gaining increasing importance across various fields such as healthcare \cite{Hasselgren2020}, supply chain management \cite{queiroz2020blockchain}, finance \cite{Patel2022}, energy \cite{Bao2021}, voting systems \cite{Jafar2021} and more \cite{Zheng2018}. As the significance of blockchain technology continues to expand, there is a corresponding rise in the demand for methodologies to analyze blockchain data. A crucial aspect of such methodologies is the development of algorithms that facilitate data visualization to enable users to discern underlying patterns and structures with greater clarity. In this report, we discuss an approach to visualize so-called transaction graphs that typically arise in the context of crypto currency data. Here, in contrast to other approaches, we put special focus on the efficiency of our method w.r.t. the number of nodes and edges in the graph to handle the massive amount of transaction data.
It is worth noting that the algorithm proposed herein is not limited to transaction graphs but can be employed in scenarios where large graphs require visualization. Such scenarios may also occur within the domains mentioned earlier.

\section{Bitcoin Transaction Graphs}
\begin{figure}
    \centering
    \includegraphics[width=\textwidth]{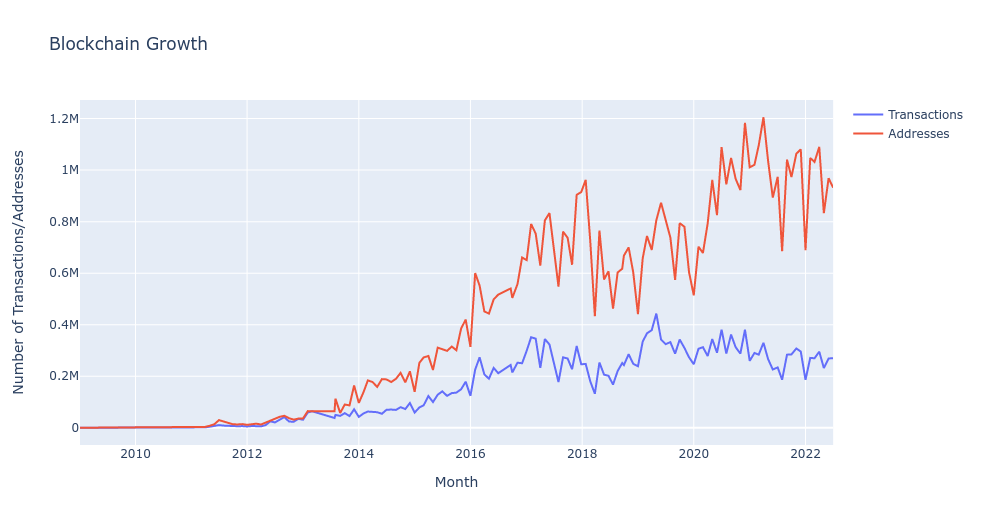}
    \caption{Number of bitcoin transactions and addresses on a monthly basis.}
    \label{fig:bitcoin_transaction_history}
\end{figure}
Bitcoin is a digital currency that has experienced significant growth and popularity since it was launched in 2009. The number of bitcoin transactions and new addresses has significantly increased over time, as we can see in Figure~\ref{fig:bitcoin_transaction_history}, which shows the history of the number of transactions and new addresses on a monthly basis. Bitcoin is not issued by any central organization. Instead, it operates with a public ledger, also known as the blockchain \cite{nakamoto2008bitcoin}. Therefore, bitcoin, among other crypto currencies, provides a great framework for studying transaction behavior. The transaction graphs that arise typically become very large and, thus, hard to process. Bitcoin processes around seven transactions per second, which sums up to around half a million transactions a day. Even worse, each transaction can contain many participating entities.

\begin{figure}
    \centering
    \includegraphics[width=0.7\textwidth]{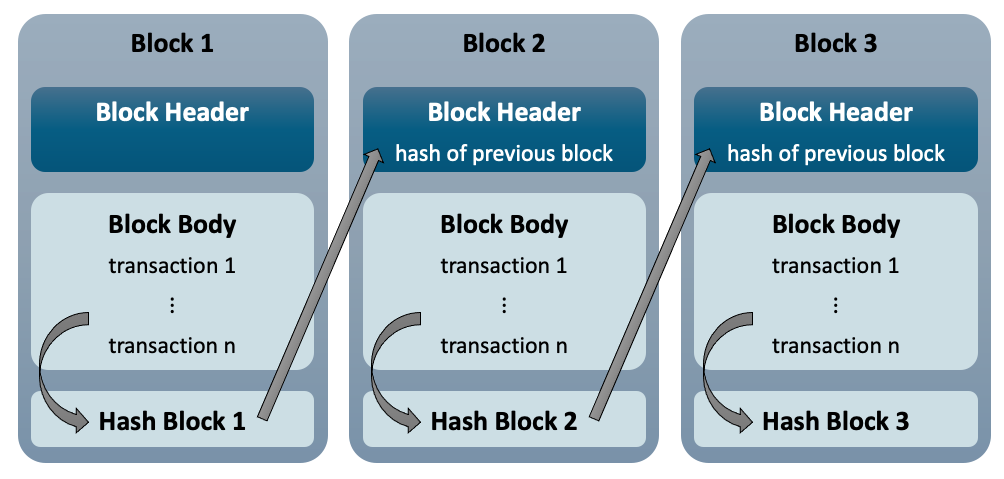}
    \caption{Schematic illustration of a blockchain.}
    \label{fig:Blockchain}
\end{figure}

The bitcoin transaction data is encoded in the blockchain. The blockchain consists of many blocks organized in a linear ordering over time, see Figure~\ref{fig:Blockchain} for a schematic illustration of the chain.

Each block has two main parts: the header section and the list of transactions. The header section contains general information about the block, such as the time it was created and a reference to the previous block. The list of transactions, on the other hand, is composed of inputs and outputs. Inputs refer to entities that send value, while outputs refer to entities that receive value. Each output contains a value to be received and a script that must be solved to authorize the spending of the value. Each input, on the other hand, consists of a hash of a previous transaction and a script that solves the problem to one of the outputs, thereby authorizing the spending of its value. It is worth noting that no value is required as an input since it spends the entire amount of a previous output.

In most cases, both the input and output scripts follow one of a couple of standardized formats, where the output problems, containing a public key, can easily be solved with the knowledge of a private key. The public keys can be interpreted as addresses. From this, we can build a first transaction graph, the vertices being the addresses and the transactions being directed hyperedges.

Working with hyperedges, albeit structurally encoded into the bitcoin transaction format, is not very practical. Following \cite{Fleder2015}, we will define the bitcoin transaction graph as follows.
The transaction graph is a bipartite graph composed of two sets of vertices: addresses and transactions. An edge exists between an address and a transaction if the former served as an input in the transaction. Conversely, an edge exists between a transaction and an address if the latter was an output in the transaction.

Note that every user usually corresponds to multiple addresses in this resulting graph. The work of \cite{zhang2020heuristic} suggests to cluster addresses used as inputs to the same transaction. Another heuristic in \cite{zhang2020heuristic} tries to abuse the fact that, by definition, the input address always spends all of its value in a transaction, so a user wanting to spend some partial value of an address might create a new address to receive the remaining value of the input address.

Better clustering is closely related to the anonymity of the blockchain \cite{Reid2012}, \cite{Goldfeder2017}, \cite{ShenTu2015}.

Our goal is to visualize this bitcoin transaction graph in certain time frames, for example, the visualization of transactions of one block or one day.

\section{Force-Directed Algorithms}
In this section, we describe the two force-directed algorithms we use as building blocks for the method described in this report. Force-directed algorithms are inspired by physical particle systems where the minimization of a certain energy functional w.r.t. the node locations leads to the layout. Here, the energy functional is induced by one or multiple forces defined between the nodes of the graph.

\subsection{Algorithm from Kamada and Kawai}
In this section, we describe the algorithm from Kamada and Kawai \cite{Kamada1989} to visualize graphs. The algorithm computes coordinates $p_i\in\mathbb{R}^2$ for each vertex $v_i$, $1\leq i\leq N$, in the graph by minimizing the \emph{energy norm} 
\[
\sum_{i=1}^N\sum_{j=i+1}^N \frac{1}{d_{ij}^2}\left(|p_i-p_j|-l_{ij}\right)^2.
\]
Here, $d_{ij}$ denotes the length of the shortest path between vertex $i$ and $j$ and $l_{ij}:=l\cdot d_{ij}$ for some rescaling constant $l>0$.
To minimize the energy, we use the Newton-Raphson method. 
Note that to find the shortest path between each pair of vertices, we use the Floyd–Warshall algorithm \cite{Floyd1962}, which has a runtime of $O(|V|^3)$ and needs $O(|V|^2)$ storage for the pairwise distances. One may further reduce the runtime for our sparse transactions graphs to $O(|V|^2\log |V| + |E||V|)$ by using Johnson's algorithm \cite{Johnson1977}. However, in both cases, typical daily bitcoin transaction graphs with the number of nodes in the range of hundreds of thousands or even millions are too large to be directly handled by this approach.

One can further speed up this algorithm by clustering portions of the graph to reduce the number of vertices, rendering the layout, and finally, declustering. Examples of such clustering/hierarchical methods are:
\begin{enumerate}
    \item Contracting nodes with only one edge to their neighbours. Then declustering can be done by locating this node using the given distance of the edge to the neighbour away from the center of gravity of the layout.
    \item Removing nodes with only two edges and replacing both edges with a single edge. To decluster, simply replace the new edges with two edges with a node in the center.
\end{enumerate}
Compare also \cite{Kobourov2016} for a discussion about handling large graphs to speed up the algorithm. However, we apply a different method in the case of components with many edges or vertices that lead to good results in a lot of examples, which we will describe in the following section.

\subsection{ForceAtlas2 with Fast Multipole}
The approach presented here belongs to a line of approaches simulating a physical system where nodes repulse each other and edges attract the incident nodes \cite{Fruchterman1991, Martin2011, Yifan2005}. It is closely related to the ForceAtlas2 algorithm \cite{Jacomy2014} while we are replacing the Barnes-Hut algorithm with the Fast Multipole method, see subsection \ref{sec:FastMultipole}.

In each iteration, we calculate several forces that apply to the different nodes. For each edge, we calculate an attraction force on the incident nodes, which is proportional to the distance between incident nodes. We further have a gravity force pulling all nodes to the center, proportional to the distance to the center. Finally, we have repulsion forces between all node pairs proportional to one over their distance.

The next step is to apply the calculated forces to each node. We follow \cite{Jacomy2014} on how to choose the step size. For this, in each iteration $t$, we define two values, the swing: 
$$swg = \sum_n |F_t(n)-F_{t-1}(n)|$$
and traction:
$$trc = \sum_n |F_t(n)+F_{t-1}(n)|$$
where $F_t(n)$ are the aggregated forces of a node $n$ in iteration $t$. Note that a big swing signals a large variance in the forces between iterations and, therefore, much more erratic movement. On the other hand, if the traction is large, this signals progress in some sense as we continue to push nodes in the same direction. Now, we iteratively adjust the step size to keep the ratio $\frac{trc}{swg}$ in some tolerated interval.

\subsubsection{Fast Multipole Algorithm}\label{sec:FastMultipole}
In each iteration of the force-directed algorithm, the forces on each node need to be calculated. As the bitcoin transaction graphs are not very dense, we can efficiently calculate the forces generated by the edge attraction and also the gravity attraction. However, computing the repulsion forces between each pair of nodes in a straightforward way is of quadratic complexity in the number of nodes, which may be very time-consuming considering the number of nodes of a typical transaction graph. Here, the Barnes-Hut algorithm \cite{Barnes1986}, developed initially to speed up the computation for physical systems, reducing the complexity to $O(N\log(N))$, has been successfully applied in the context of graph visualization, see \cite{Jacomy2014}. However, we can further improve complexity to $O(N)$ by using the Fast Multipole algorithm \cite{Greengard1987}, which we will briefly describe in the following.

In a first step, the algorithm computes a quad tree structure of the given particles. Here, we start with a square that covers all the particles. If a square contains more than a certain number of particles, we divide it into four new squares. Each of these squares is called a cell. This generates a tree structure of cells, a cell being the parent of another if it was generated by one subdivision of the parent. For each cell, we define the neighbouring cells to be those adjacent cells that have minimal size but are not smaller than the original cell. Further, we define interacting cells of a cell to be the minimal cells that are neighbouring the parent of the cell or are children of a cell neighbouring the parent.

The trick of the Fast Multipole method is to calculate two Taylor expansions of a chosen degree for each cell around its center. The first approximation is for the force that nodes act on particles that are far away, called the outgoing expansion. The second approximation is for the force that nodes in the cell feel from particles that are far away, called the incoming expansion. Here, the term ``far away'' refers to the nodes not located in neighbouring cells.

The idea is to calculate these expansions recursively. The recursive calculation of the expansions is performed by first computing the outgoing expansions. The method starts by computing the outgoing expansion for each leaf cell. Subsequently, the quad tree is traversed upwards, and the outgoing expansions of the four children of a cell are used to calculate its outgoing expansion. This is accomplished by shifting the center of the children's outgoing expansions to the center of the new cell and adding them up. The quad tree is then traversed in reverse order to calculate the incoming expansions. This is accomplished by applying a center shift of the incoming expansion of the parent and adding all outgoing expansions of the interaction neighbours reformulated as incoming expansions of the cell to it.

Finally, we can calculate the force that applies to each single node. For each node, let the corresponding leaf cell be the minimal cell containing the node. Now, for every node, we apply the incoming expansion of its leaf cell to the node to get the force from far away nodes and additionally add all the forces from nodes in neighbouring cells of the leaf cell and the leaf cell itself to the node.

\section{Final Algorithm}
\begin{figure}
    \centering
    \includegraphics[width=\textwidth]{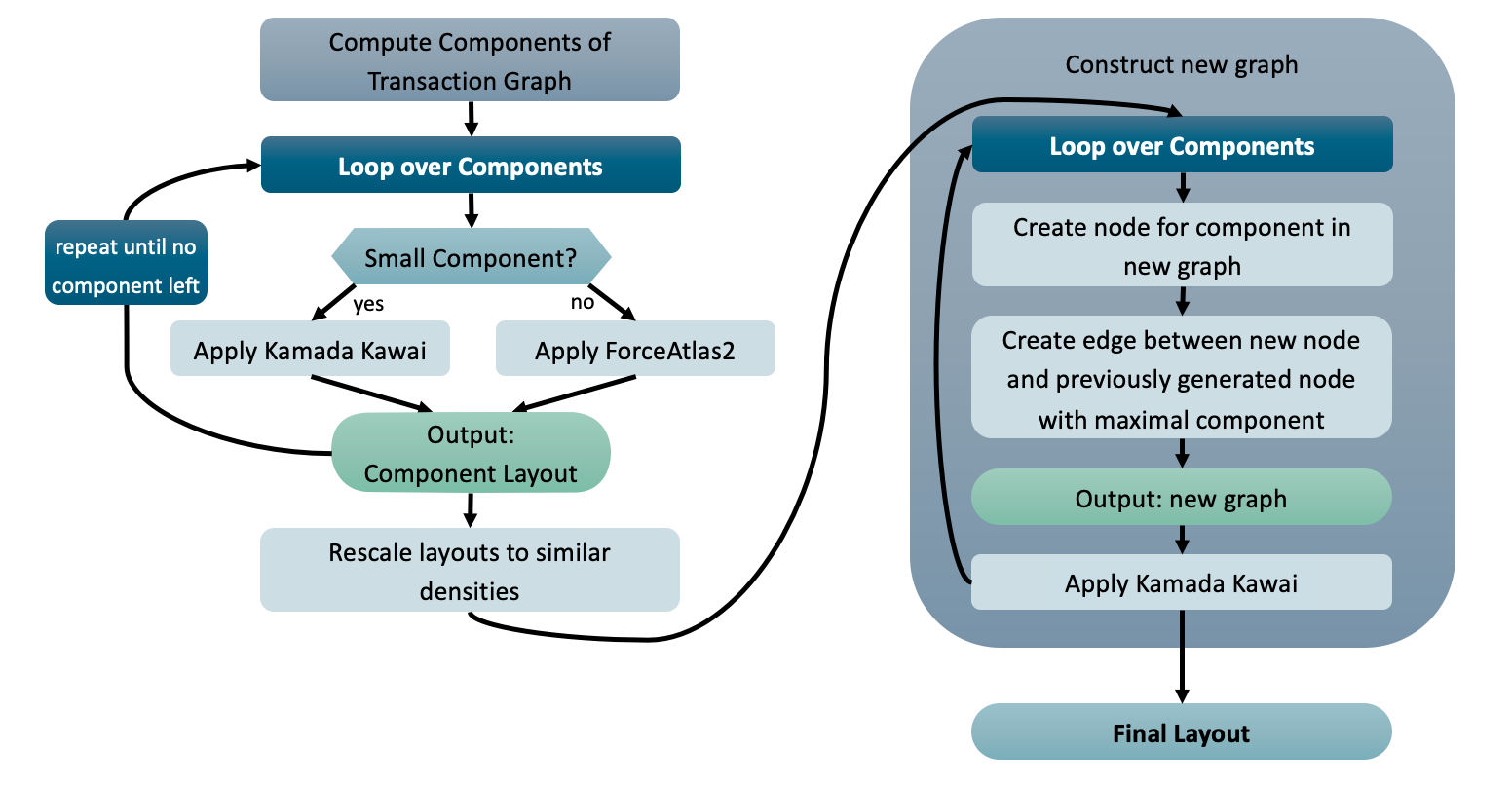}
    \caption{Illustration of the overall algorithm for blockchain data visualization.}
    \label{fig:Algorithm}
\end{figure}
The final algorithm that we apply to visualize the transaction data now combines the algorithms above to optimize the tradeoff between quality and computational time. 

Here, we first split the given graph into its components (ignoring the directions of the edges). Note that the typical transaction graph for a certain time period most often consists of many components, where most of the components consist of only a few nodes.
Now, for each component, we calculate a separate layout. In cases with a few nodes, we opt for the Kamada-Kawai algorithm, which typically generates more expressive layouts. For components containing a high number of nodes, we employ the force-directed algorithm based on the Fast Multipole method described above.

These component layouts are then rescaled to have a similar density. Next, we assemble the separate component layouts into a single figure. To accomplish this, we construct a new graph with the components represented as nodes, connecting them with edges to form a tree structure.
These edges are determined by selecting a random order and sequentially linking the next component to the largest previous component. We choose the edge sizes to be half of the diameters of both components summed up plus some constant. If this resulting graph is small enough, we apply the Kamada-Kawai algorithm. Otherwise, we divide it into several components and apply the Kamada-Kawai algorithm individually to each component and reassemble them. Figure~\ref{fig:Algorithm} shows an illustration of the overall algorithm.

\begin{figure}[H]
    \centering
    \includegraphics[width=\textwidth]{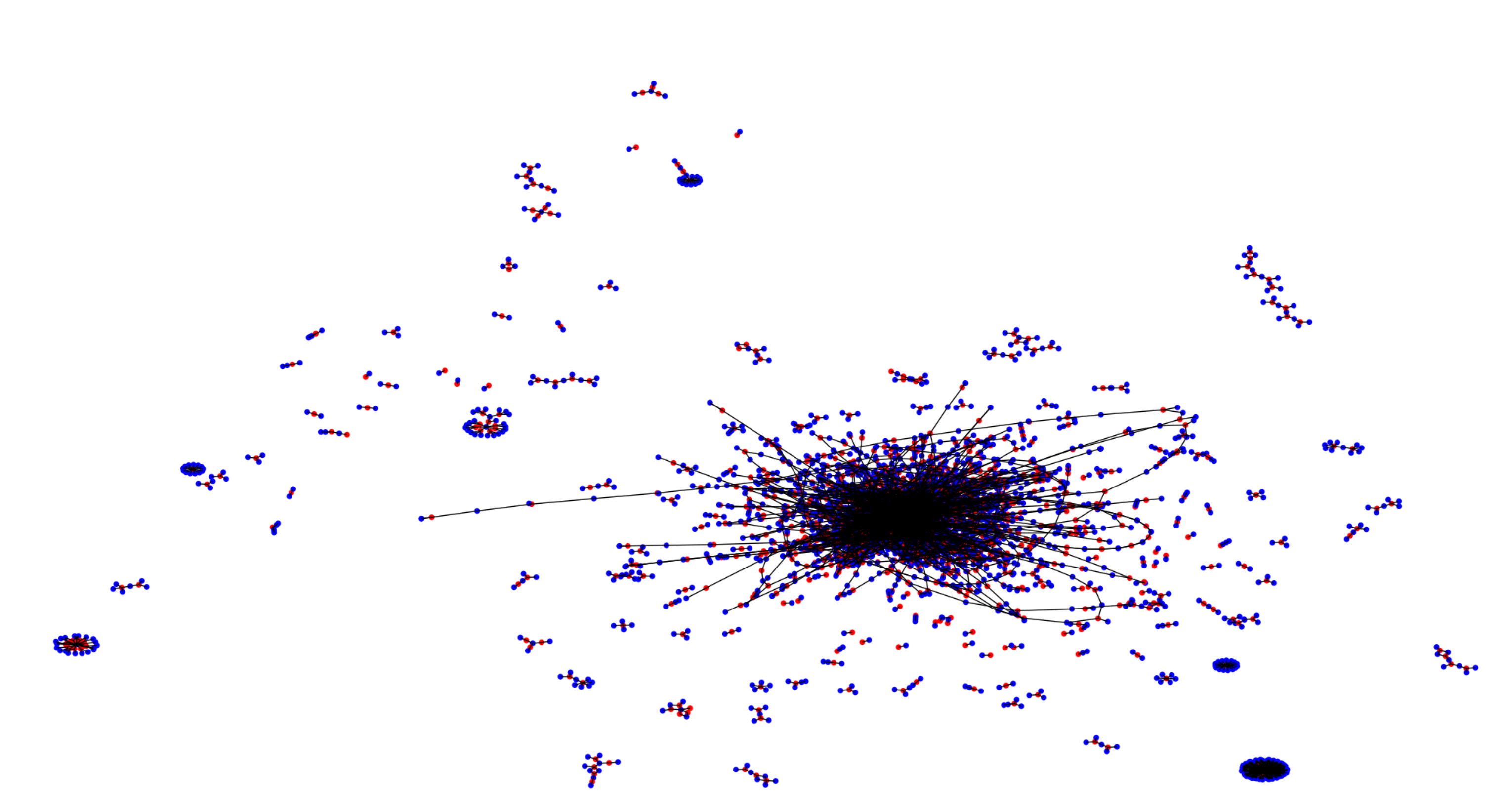}
    \caption{A visualization of the transaction graph on April 19, 2011. Transaction nodes are red, address nodes are blue. Contains around $1.5\cdot10^4$ nodes.}
    \label{fig:19411}
\end{figure}

\begin{figure}[H]
    \centering
    \includegraphics[width=\textwidth]{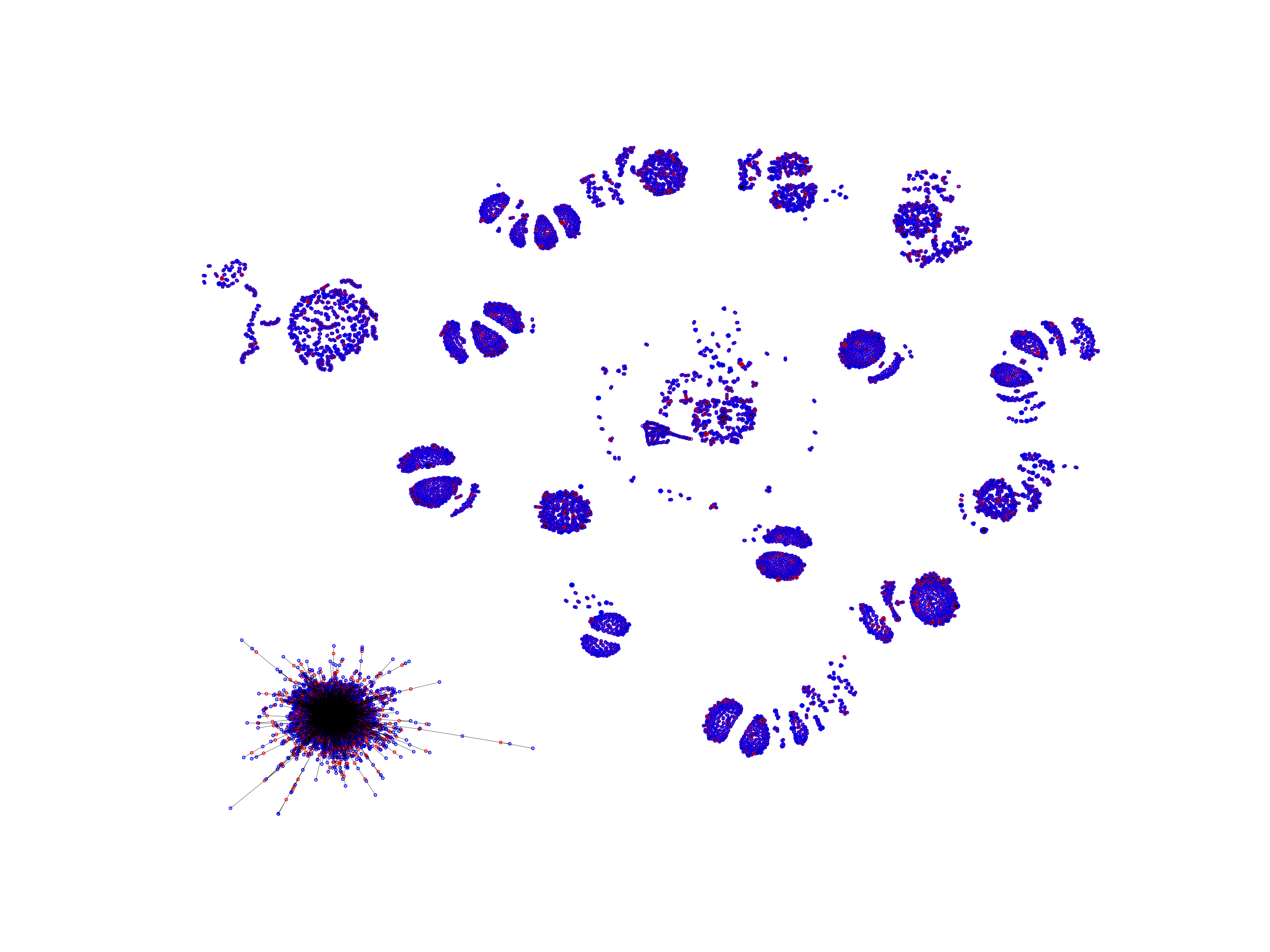}
    \caption{A visualization of the transaction graph on December 12, 2013. Transaction nodes are red, address nodes are blue. Contains around $2\cdot10^5$ nodes.}
    \label{fig:121213}
\end{figure}

Figures~\ref{fig:19411} and~\ref{fig:121213} show the resulting visualization of the transaction graphs at two different points in time.

\section{Summary}
In this report we described an algorithm to efficiently visualize graphs with a very high number of nodes and edges. In our experiments, this algorithm produced meaningful graphs within a reasonable amount of computing time, while other algorithms were not applicable due to their computational complexity.

Although developed for analyzing bitcoin blockchain data, it may be worth applying this algorithm to other blockchain transaction graphs or large graphs from other domains.

We provide the source code of this algorithm in the public GitHub repository at \url{https://github.com/frontmark/research}.

\newpage
\bibliographystyle{plain}
\bibliography{literature}

\end{document}